\documentclass[aps,showpacs,twocolumn,amsmath,amssymb,prl]{revtex4-1} 

\usepackage{graphicx} 
\usepackage{bm} 

\begin{document} 
 
\title{
Crossover dynamics of dispersive shocks in Bose-Einstein condensates characterized by two and three-body interactions
} 
 
\author{M. Crosta$^{1,3}$, S. Trillo$^{2}$ and A. Fratalocchi$^{1}$} 

\email{andrea.fratalocchi@kaust.edu.sa} 
\homepage{www.primalight.org}

\affiliation{
$^1$PRIMALIGHT, Faculty of Electrical Engineering; Applied Mathematics and Computational Science, 
King Abdullah University of Science and Technology (KAUST), Thuwal 23955-6900, Saudi Arabia\\
$^2$Dipartimento di Ingegneria, Universit\`{a} di Ferrara, Via Saragat 1, 44122 Ferrara, Italy\\
$^3$Dept. of Physics, Sapienza University of Rome, I-00185, Rome, Italy 
} 

\date{\today} 
 
\begin{abstract} 
We show that the perturbative nonlinearity associated with three-atom interactions, competing with standard two-body repulsive interactions,
can change dramatically the evolution of 1D dispersive shock waves in a Bose-Einstein condensate.
In particular, we prove the existence of a rich crossover dynamics, ranging from the formation of multiple shocks regularized by coexisting trains of dark and antidark matter waves, to 1D soliton collapse. For a given scattering length, all these different regimes can be accessed by varying the number of atoms in the condensate.
\end{abstract} 
 
 
\maketitle 
\paragraph{Introduction. ---}
Bose-Einstein Condensation (BEC) has been successfully described in the mean-field limit
by the Gross-Pitaevskii  equation (GPE), which embodies quantum two-body interactions (\emph{s}-wave scattering) 
into its cubic nonlinear term \cite{DalfovoRMP99}. Though not compulsory for (ideal, i.e. interaction-less) condensation,
such term sustains several, remakable and otherwise unobservable, coherent phenomena such as soliton formation \cite{Becker08},
quantum shocks \cite{Dutton01,Hoefer06}, collapse \cite{collapse}, nonclassical states \cite{nonclass}, 
time reversal, quantum chaos and Anderson localization \cite{treversal}. 
The exceptionally huge tunability (in magnitude and sign) of the scattering length $a$ through Feshbach resonances \cite{RevModPhys.82.1225}, have made the area of ultracold atoms a prolific ground to investigate new quantum and nonlinear physics.
Particular interest, in this context, was stirred by higher-order few-body interactions \cite{KohlerPRL02} boosted also by the observation of resonant Efimov states \cite{expEfimov}. Most efforts in this area are presently aimed at assessing the impact of few-body recombination loss coefficients which limits the lifetime of condensates \cite{loss3body}. 
However, almost unexplored is also the effect of three-body elastic collisions, which naturally arises as a quintic conservative (Hamiltonian) term in the higher-order expansion of the GPE  \cite{KohlerPRL02,Gammal00,Cardoso11} whose impact clearly grows as higher densities are being reached in experiments \cite{hiN_2011}. It is therefore of paramount importance to assess whether the three-body interactions can lead to a clear signature in terms of qualitative new scenarios, especially if the latter becomes accessible in the perturbative limit.\\
In this letter we address this question with reference to the dynamics of dispersive shock waves (DSWs),
which form in a {\em repulsive} BEC when the kinetic spreading regularizes the tendency driven by the nonlinearity
to form multivalued wave fronts.
At variance with earlier \cite{Dutton01,Hoefer06,Damski04} and more recent studies on DSWs
\cite{ChangPRL08,LeszczyszynPRA09,MeppelinkPRA09,DekelPRA10}, which are all based on the standard GPE,
we investigate the case characterized by an attractive three-body nonlinear correction to the repulsive \emph{s}-wave scattering.
This regime is expected to be achievable by exploiting resonance tuning for bosons and turns out to be relevant also for superfluid fermions \cite{PieriPRL03}. Unlike the case of competing nonlinearities of the same sign (where three-body contributions merely strengthen the nonlinearity at high densities), we find that the presence of the quintic term of different sign leads to a rich phase-diagram of the wave-breaking phenomenon, with novel regimes that uniquely identify the contribution of the three-body interaction. Remarkably, for a given scattering length, the crossover between different dynamics can be totally controlled by changing the number of atoms in the condensate. Due to the ubiquitous nature of DSW, these findings are also of fundamental interest in other areas of classical physics, including nonlinear optics \cite{optics},
electronic systems \cite{Bett06}, and granular chains \cite{granular09}.
\paragraph{Model and general shock scenario. --} 
We start from the following dimensionless cubic-quintic GPE, describing the free evolution
in 1D (after trap is released along $x$) of a BEC ruled by two and three-body interactions:
\begin{equation}
\label{gpsta}
i\epsilon\frac{\partial\psi}{\partial t}+\frac{\epsilon^2}{2}\frac{\partial^2\psi}{\partial x^2}-|\psi|^2\psi+\alpha |\psi|^4\psi=0,
\end{equation}
having introduced dimensionless spatial $x=\frac{X}{\epsilon}\sqrt{mN\omega/\hbar}$ and temporal $t=TN\omega/\epsilon$ coordinates, with $\epsilon=\hbar/m\omega R_{TF}\ll 1$ ($R_{TF}$ is the Thomas-Fermi radius), and a rescaled wave function $\psi=\Psi\sqrt{4\pi\hbar a /m\omega N}$, being $m$ the single boson mass, $N=\int|\Psi|^2dX$ the number of particles and $\omega$ the transverse BEC trap frequency.
\begin{figure}
\centering
\includegraphics[width=9cm]{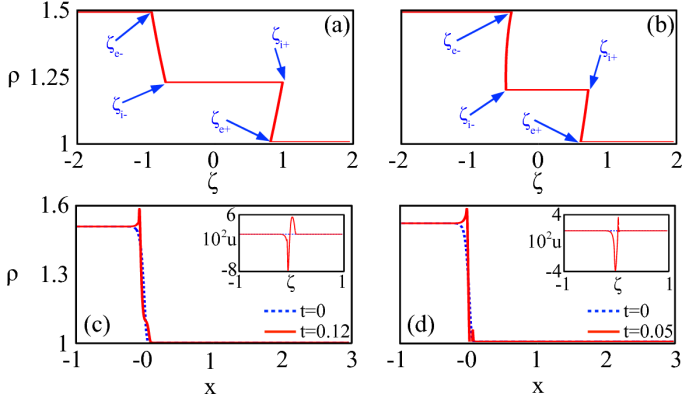}
\caption{Color Online. Dispersionless evolution for $\rho_0^+=1.5$: (a-b) as obtained from Eqs. (5-7) with $\alpha=0.15$ (a), $\alpha=0.3$ (b);
(c-d) numerical results from Eqs. (2) comparing the case $\alpha=0.4$ (c) with the case of a standard GPE ($\alpha=0$) with {\em positive} cubic nonlinearity (d). 
\label{sha}
}
\end{figure}
The parameter $\alpha/\hbar\omega=N g_3/|g_2|^2$ (with $\alpha\ge 0$) uniquely determines the properties of the system in terms of the number of particles and the relative strength of the three-body term $g_3$ \cite{KohlerPRL02} over the two-body term $g_2=4\pi\hbar^2 a/m$. At variance with the cubic GPE system ($\alpha=0$), which is completely characterized by the scattering length $a$, this normalization clearly shows that the high-order BEC dynamics lives in a two dimensional phase space encompassing both $a$ and $N$. This will permit to explore different scenarios even for a constant scattering length. We study the general shock dynamics in the dispersionless (or hydrodynamic) limit of Eq. (\ref{gpsta}), which can be obtained by applying the Madelung transformation $\psi=\sqrt{\rho}\exp[\frac{i}{\epsilon}\int dx u]$ for $\epsilon\rightarrow 0$:
\begin{align}
\label{dlim}
&\frac{\partial\rho}{\partial t}+\frac{\partial(u\rho)}{\partial x}=0, &\frac{\partial u}{\partial t}+u\frac{\partial u}{\partial x}+\frac{\partial f}{\partial x}=0,
\end{align}
where $f(\rho)=\rho-\alpha\rho^2$, and $\rho$, $u$ act as density and velocity of the fluid, respectively. 
The impact of the quintic term can be understood by studying the decay of an initial jump in the density,
which can be easily realized experimentally. This amounts to solve the Riemann problem associated with
the initial value $u(x,0)=0$, $\rho(x,0)=\rho_0^+ + (\rho_- - \rho_0^+)\Theta(x)$ [$\Theta(x)$ is the Heaviside function], 
$\rho_0^{\pm}$ ($\rho_0^{-}>\rho_0^{+}$) being the initial constant densities across $x=0$.
We solve Eqs. (\ref{dlim}) by deriving the following Riemann invariants $\lambda_{\pm}=u\pm R(\rho)$:
\begin{align}
\label{diag}
R(\rho)=\sqrt{\rho(1-2\alpha\rho)}+\frac{\cos^{-1}(1-4\alpha\rho)}{2\sqrt{2\alpha}},
\end{align}
which transform Eqs. (\ref{dlim}) into the diagonal form $\frac{\partial\lambda_{\pm}}{\partial t}+V_{\pm}\frac{\partial\lambda_{\pm}}{\partial x}=0,$
where $V_{\pm}=u\pm\sqrt{\rho(1-2\alpha\rho)}$ are the Riemann characteristic eigenvelocities. 
As long as $V_\pm$ are strictly real, i.e.
\begin{equation}
\label{cond}
\alpha<1/(2\rho_0^-),
\end{equation}
the evolution of $\lambda_\pm$ can be given in terms of the self-similar variable $\zeta=x/t$
through the following simple wave solutions of the system (characterized by one of the two Riemann variables remaining constant) 
\begin{align}
\label{sol}
&R(\rho_0^+)-R(\rho)-\sqrt{\rho(1-2\alpha\rho)}=\zeta, \nonumber\\
&R(\rho)-R(\rho_0^-)+\sqrt{\rho(1-2\alpha\rho)}=\zeta.
\end{align}
The two simple waves (\ref{sol}) are connected by an intermediate region with constant values $\rho=\rho_i$ and $u=u_i$.
The latter can be found by the matching equations $u_i=[R(\rho_0^+)-R(\rho_0^-)]/2$ and $2R(\rho_i)=R(\rho_0^+)+R(\rho_0^-)$. 
The borders $x_{i\pm}$ of the intermediate region are then calculated by substituting the density $\rho_i$ into Eqs. (\ref{sol}):
\begin{align}
\label{pts}
\frac{x_{i\pm}}{t}\equiv\zeta_{i\pm}=R(\rho_i)-R(\rho_0^-)\pm\sqrt{\rho_i (1-2\alpha\rho_i)}.
\end{align}
Conversely, the simple wave external edges $x_{e\pm}\equiv\zeta_{e\pm}t$ 
are calculated by matching Eqs. (\ref{sol}) with the asymptotic (input) values $\rho_0^{\pm}$, thus obtaining
\begin{align}
\label{pts1}
&\frac{x_{e-}}{t}=-\sqrt{\rho_0^+(1-2\alpha\rho_0^+)}, &\frac{x_{e+}}{t}=\sqrt{1-2\alpha}.
\end{align}
Equations (\ref{sol})-(\ref{pts1}) predict a rich scenario for the shock dynamics, as illustrated in Fig. \ref{sha}a-b.
Henceforth, without loss of generality, we take $\rho_0^+=1$ and $\rho_0 \equiv \rho_0^-$. 
For small values of $\alpha$ [see Fig. \ref{sha}a], wave-breaking occurs only for $x>0$, where the velocity $\zeta_{i+}$ of the point $x_{i+}$ is larger than $\zeta_{e+}$, leading to a multivalued region, whose dispersive regularization leads to a  DSW.
\begin{figure}
\centering
\includegraphics[width=8cm]{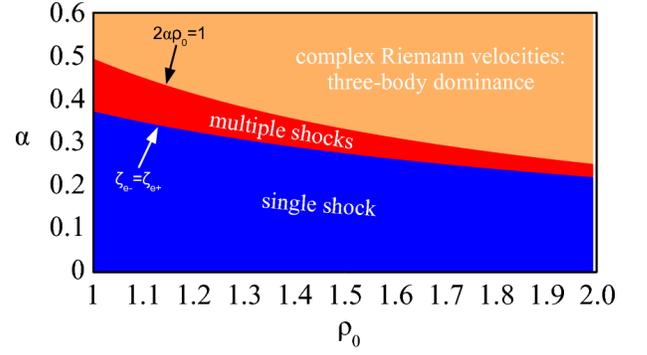}
\caption{Color Online. Global phase diagram of the shock dynamics
in the parameter plane $(\rho_0 \equiv \rho_0^-, \alpha)$ [$\rho_0^+=1$]. 
\label{pha}
}
\end{figure}
However, a more complex dynamics is observed when $\alpha$ is increased [see Fig. \ref{sha}b]: in this case not only $x_{i+}$ but also $x_{e-}$ moves sufficiently fast and induces wave breaking for $x<0$, thus producing a double shock in the dynamics. The crossover between these two regimes is given by the condition $\zeta_{e-}(\alpha^*)=\zeta_{i-}(\alpha^*)$. It is worth noting that a double shock have also been predicted in a BEC flowing through a penetrable barrier \cite{LeszczyszynPRA09}, though both the underlying mechanism and the regularization are completely different in the latter case.
A third regime settles in when Eq. (\ref{cond}) is violated, resulting in imaginary eigenvelocities $V_{\pm}$ [Eqs. (\ref{dlim}) are no longer hyperbolic]. In this case, solutions (\ref{sol})-(\ref{pts1}) are meaningless and we resort to numerical integration of Eqs. (\ref{dlim}). Our analysis show (Fig. \ref{sha}c) that the system evolution is characterized by the generation of two opposite velocities (Fig. \ref{sha}c, $u$ in the inset), which compress the wavefront and generate a cusp-like singularity at $x\approx 0^-$. This behavior originates from a three-body dominance over the two-body term, as demonstrated by comparing Fig. \ref{sha}c to Fig. \ref{sha}d, which shows the evolution of the same input launched in a standard GPE with positive nonlinearity [Eq. (\ref{gpsta}) with $\alpha=0$ and $x\rightarrow ix$]. The two wave-breaking are qualitatively identical,  which can be understood by considering that the GPE with attractive two-body nonlinearity exhibits complex eigenvelocities, in the same way as Eq. ({\ref{gpsta}}) for $\alpha >1/2\rho_0$. Despite  such similarity, the long-term dynamics (regularization) in the presence of three-body nonlinearities will be totally different, as discussed in the following paragraphs. Figure \ref{pha} summarizes the results of the dispersionless analysis in a wave-breaking phase diagram. In particular, the two curves $\alpha^*=\alpha^*(\rho_0)$ and $\alpha=1/(2\rho_0)$, divide the phase space into three distinct regions, each characterized by different wave breaking scenarios deepened below. 
\begin{figure}
\centering
\includegraphics[width=8.5cm]{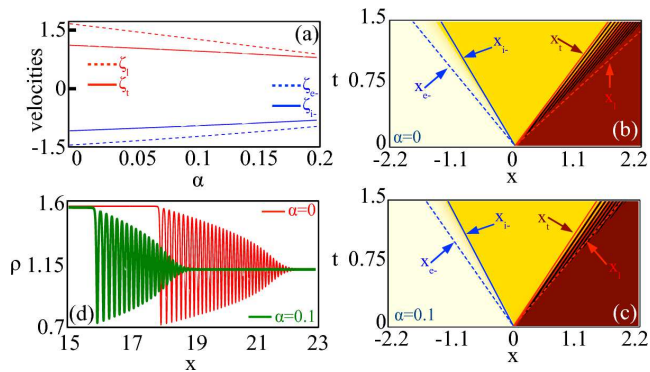}
\caption{Color Online.  (a) Velocity of DSW edges $\zeta_t$, $\zeta_l$, and rarefaction edges $\zeta_{e-}$, $\zeta_{i-}$ vs. $\alpha$; 
(b-c) Color level plot of density evolution $\rho(x,t)$ from Eq. (1) for $\epsilon=0.01$ and (b) $\alpha=0$, (c) $\alpha=0.1$; 
For comparison the edges of the DSW $x_{l}$, $x_{t}$ from Eq. (\ref{velg}), 
$x_{e-}$, $x_{i-}$ of the rarefaction wave from Eqs. (\ref{pts})-(\ref{pts1}) are reported. 
(d) Snapshots $\rho(x,t=8)$ from (b-c) showing the DSW dark soliton train for $\alpha=0,0.1$.  
\label{num1}
}
\end{figure}
\paragraph{Real eigenvelocities: tuning the shock dynamics. --}
We study the shock regularization for $\alpha\le\alpha^*$ by exploiting Whitham theory of modulation, in the form suitable
for nonintegrable systems \cite{El05}. We summarize only the outcome of this approach, while deferring the (involved) mathematical details to a successive paper. When $\alpha\le\alpha^*$, the multivalued region for $x>0$ (Fig. \ref{sha}) is regularized by the formation of a single DSW such that a modulated cnoidal wave (or dark soliton train) appears within a shock fan limited by a leading $x_l$ and a trailing $x_t$ edge.
By matching the dispersionless limit with the Whitham equations \cite{El05}, we are able to find the velocities of the edges $\zeta_l=x_l/t$ and $\zeta_t=x_t/t$ (replacing the hydrodynamic estimates $\zeta_{e+}$, $\zeta_{i+}$), in the form:
\begin{align}
\label{vels}
&\zeta_l=(1-2\alpha)\bigg(2\gamma_l-\frac{1}{\gamma_l}\bigg),\nonumber\\ 
&\zeta_t=R(\rho_i)-R(1)+\gamma_t\sqrt{\rho_i (1-2\alpha\rho_i)},
\end{align} 
with $\gamma_l=\gamma(1)$ and $\gamma_t=\gamma(\rho_i)$ arising from the solution of the differential equation:
\begin{align}
\label{velg}
\frac{\partial\gamma}{\partial\rho}=\frac{(1+\gamma)(1+2\gamma-8\gamma\alpha)}{2\rho(2\gamma+1)(2\alpha\rho-1)},
\end{align}
integrated in $\rho\in[1,\rho_i]$, with the initial condition $\gamma(1)=1$ for $\gamma_l$, and $\gamma(\rho_i)=1$ for $\gamma_t$. Figure \ref{num1}a displays the DSW edge velocities $\zeta_l$, $\zeta_t$ [calculated by integrating Eq. (\ref{velg})] and those of the rarefaction wave $\zeta_{e-}$, $\zeta_{i-}$  [from Eqs. (\ref{pts})-(\ref{pts1})] as a function of $\alpha$. Increasing the three-body contributions results into a nearly linear decrease of all relevant velocities $\zeta_l$, $\zeta_t$, $\zeta_{e-}$, $\zeta_{i-}$, with a consequent reduction of the extension of both simple waves and the shock fan. It is worth emphasizing that the two edges of the DSW behave in a markedly different way: the leading edge $\zeta_l$ is strongly influenced by $\alpha$, while the variation of $\zeta_t$ is much smaller. As a consequence, even when $\alpha$ is increased by a small factor, the shock dynamics is significantly affected in terms of both shock fan and shock overall angular direction. To verify these predictions, we resort to numerical integration of Eq. (\ref{gpsta}). As shown in Fig. \ref{num1}b-c, a remarkable agreement is obtained between theory and numerical simulations. The snapshot in Fig. \ref{num1}d shows that, despite a small variation of $\alpha$, a substantial reduction of the shock fan extension occurs (nearly a factor of two for $\Delta \alpha=0.1$)
accompanied by a reduced average velocity (center of mass closer to $x=0$). In summary, in the regime where a single shock is formed for $x>0$, all the features of the shock dynamics are significantly varied by slightly perturbing the strength of the three-body interaction terms $\alpha$.
\begin{figure}
\centering
\includegraphics[width=9cm]{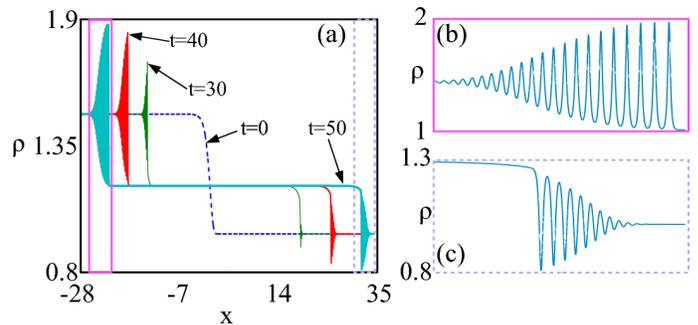}
\caption{Color Online. (a) Snapshots of density $\rho(x)$ obtained from Eq. (1) with $\alpha=0.3$, $\epsilon=0.01$, and a input jump $\rho_0=1.5$; 
(b)-(c) zoom of the boxed evolution in (a), showing the opposite type of regularization taking place in the dynamics.  
\label{num2}
}
\end{figure}
\paragraph{Real eigenvelocities: multi-shock generation through antidark solitons. --}
When $\alpha^* <\alpha< 1/2\rho_0$, simulations show the occurrence of a second DSW for $x<0$ (Fig. \ref{num2}), in complete agreement with the hydrodynamic analysis. In this case, however, the DSW for $x<0$ differs substantially from the dynamics discussed above for $x>0$. The latter, in fact, is characterized by a cnoidal wave composed by a train of dark-like oscillations (see Fig. \ref{num1}), owing to the condition $\rho(x_t)>\rho(x_l)$ (see also Fig. \ref{sha}a). The DSW, more specifically, originates from a series of oscillations that start from the leading edge $x_l$ (where the shock matches the linear wave background at $\rho=1$) and culminate into a dark soliton at the trailing edge [Fig. \ref{num2}c]. On the contrary, the DSW for $x<0$ moves in the opposite (backward) direction and has a larger density on the leading edge, which yields $\rho(x_{l1})>\rho(x_{t1})$, being $x_{l1}$ and $x_{t1}$ the leading and trailing edge of the DSW at $x<0$, respectively. As a consequence, the modulated wave train that regularizes the shock needs to be composed by bright entities, as shown in Fig. \ref{num2}b. In this case, the trailing edge is an antidark (bright on pedestal) soliton solution of Eq. (1), which has been thoroughly investigated recently \cite{CrostaPRA}. Quite remarkably, although these solitons are mostly unstable \cite{CrostaPRA}, the average velocity of the DSW for $x<0$ is such that the soliton (and hence the DSW) is totally stable on propagation.
\begin{figure}
\centering
\includegraphics[width=8.5cm]{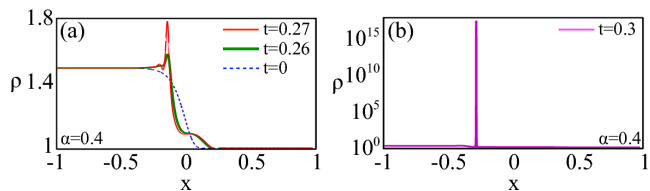}
\caption{Color Online. Snapshots of density $\rho$ vs. $x$ from Eq. (1) for $\alpha=0.4$ and $\epsilon=0.01$: 
(a) shock regularization; (b) antidark collapse instability.  
\label{num3}
}
\end{figure}
\paragraph{Complex eigenvelocities: 1D BEC collapse. ---}
When $\alpha>1/2\rho_0$, the wave-breaking observed in the simulations of Eq. (1) reflects indeed the dominant character of three-body interactions (Fig. \ref{num3}a-b). In the early stage, a cusp-like singularity in $x\sim 0^-$ is generated (Fig. \ref{num3}a), as predicted by Eqs. (\ref{dlim}). Contrary to the the case of the standard attractive GPE, which shows a similar early stage behavior [see Fig. \ref{sha}d], the three-body dominance in Eq. (\ref{gpsta}) leads to a completely different long term dynamics. In the quintic case, in fact, the singularity tends to evolve into antidark solitons, due to the condition $\rho(x_l)>\rho(x_t)$. Contrary to the previous case, however, here the solitons velocity is nearly vanishing. Zero-velocity antidark solitons of Eqs. (1) are always unstable \cite{CrostaPRA}, and they lead the BEC towards irreversible collapse near $x=0$ (Fig. \ref{num3}b). We emphasize that such a collapsing dynamics, which we have verified to occur also for different inputs (e.g. Gaussian on pedestal), is a unique features of Eq. (\ref{gpsta}). Collapse, in fact, cannot be observed in the standard GPE ($\alpha=0$) that, owing to its integrable nature, does not possess unstable solitons. It is also worthwhile remarking the perturbative value of the quintic term ($\alpha>0.34$) which induces such a dramatic and abrupt change in the dynamics.\\
In conclusion, we have shown that the perturbative nonlinearity arising from three-body
elastic collisions can dramatically alter the breaking scenario in repulsive BEC.
The generation of multiple DSWs involving antidark soliton trains, as well as 1D collapsing dynamics 
are new peculiar behaviors that can be controlled by means of the BEC number of atoms. These results are expected not only to foster new perspectives in BEC physics, but also to stimulate novel experiments in nonlinear optics, where the quintic terms account for saturation of classical Kerr nonlinearities. We acknowledge funding from PRIN 2009 project (No. 2009P3K72Z).

\end{document}